\documentclass[a4paper,fleqn]{cas-dc}

\usepackage[square,numbers]{natbib}

\usepackage{babel}

\usepackage{microtype}
\usepackage{nameref}
\usepackage{scalerel}
\usepackage{lipsum,adjustbox}
\usepackage{graphicx}
\usepackage{caption}
\usepackage{subcaption}
\usepackage{tikz}
\usepackage{tkz-graph}  

\usetikzlibrary{graphs}
\usetikzlibrary{shapes.geometric}%
\usetikzlibrary{positioning}
\usetikzlibrary{arrows.meta}

\begin{document}
\let\WriteBookmarks\relax
\def\floatpagepagefraction{1}
\def\textpagefraction{.001}

\shorttitle{Generation of Probabilistic Synthetic Data for Serious Games: A Case Study on Cyberbullying}

\shortauthors{J. Pérez}

\title [mode = title]{Generation of Probabilistic Synthetic Data for Serious Games: A Case Study on Cyberbullying}               

%

\author[1]{Jaime Pérez}[orcid=0000-0001-7511-2910]
\cormark[1]
\ead{jperezs@comillas.edu}

\author[1,2]{Mario Castro}[orcid=0000-0003-3288-6144]

\author[3]{Edmond Awad}[orcid=0000-0001-7272-7186]

\author[1]{Gregorio López}[orcid=0000-0001-9954-3504]

\affiliation[1]{organization={Institute for Research in Technology (IIT), ICAI Engineering School, Universidad Pontificia Comillas},
    city={Madrid},
    postcode={28015}, 
    country={Spain}}
\affiliation[2]{organization={Grupo Interdisciplinar de Sistemas Complejos (GISC)},
    }

\affiliation[3]{organization={Department of Economics, University of Exeter},
    city={Exeter},
    postcode={EX4 4PU}, 
    country={United Kingdom}}

\cortext[cor1]{Corresponding author}

\begin{abstract}
Synthetic data generation has been a growing area of research in recent years. However, its potential applications in serious games have not been thoroughly explored. Advances in this field could anticipate data modelling and analysis, as well as speed up the development process. To try to fill this gap in the literature, we propose a simulator architecture for generating probabilistic synthetic data for serious games based on interactive narratives. This architecture is designed to be generic and modular so that it can be used by other researchers on similar problems. To simulate the interaction of synthetic players with questions, we use a cognitive testing model based on the Item Response Theory framework. We also show how probabilistic graphical models (in particular Bayesian networks) can be used to introduce expert knowledge and external data into the simulation. Finally, we apply the proposed architecture and methods in a use case of a serious game focused on cyberbullying. We perform Bayesian inference experiments using a hierarchical model to demonstrate the identifiability and robustness of the generated data.

\end{abstract}


\begin{highlights}
\item Synthetically generated data allows anticipating data modelling and analysis, as well as speeding development process
\item We propose a simulator architecture to generate synthetic probabilistic data for serious games based on interactive narrative
\item To simulate the interaction between players and game decisions, we use a cognitive testing model (Item Response Theory)
\item We show how Probabilistic Graphical Models can be used to bring expert knowledge and external data into the simulation
\item We apply the proposed architecture and methodology in a use case of a serious game focused on cyberbullying
\end{highlights}


\maketitle

\section{Introduction}
\label{intro}

The Internet has become an integral part of young people's lives. Minors under 18 years old already accounted for nearly one-third of Internet users worldwide back in 2017, according to UNICEF's "Children in a Digital World" report \cite{UNICEFchildren2017}. The COVID-19 pandemic has enlarged such a phenomenon, making more minors spend more time online at younger ages. On the positive side, digital technology and hyper-connectivity come with significant educational and economic opportunities and better access to information.

However, uncontrolled access to the Internet also opens the door to new threats targeted toward minors, making them more accessible to bullies, harassers, and sex offenders. Children who were already vulnerable are now at greater risk. Around 10\% of European children are already victims of cyberbullying (CB) every month \cite{EUkidsOnline2020}, and 49\% have experienced a CB-related situation at least once \cite{ChidrenRisk_Covid2021}. Among European children who had already reported being a victim of CB, 44\% reported an upsurge during the 2020 COVID-19 lockdown \cite{ChidrenRisk_Covid2021}. 

Traditionally, Law Enforcement Agencies (LEAs) have focused their efforts on this problem from the criminal component. However, some recent interdisciplinary initiati\-ves (the H2020 RAYUELA project is a prominent example \cite{RAYUELA}) propose to use Serious Games (SG) as an educational and research tool, thus fostering a preventive approach. SG are explicitly designed for a primary purpose beyond pure entertainment (e.g., learning new skills, conveying values, and awareness-raising) \cite{abt1987serious}. However, being entertaining is part of their attractiveness. In this context, the RAYUELA project's central idea is to use video games' natural appeal to gather data and educate minors. Specifically, the players are immersed in an adventure where they must make decisions involving potentially hazardous cybercrime-related situations. Data will be collected to understand better (using modern data science methods) which factors influence risky online behaviour in a friendly, safe and non-invasive way.

The idea of using video games as a research tool to investigate humans is not new and has been gaining populari\-ty in recent years. For example, an experiment embedded in a video game showed that the complexity of the city where a child lives influences his or her future navigation skills \cite{Navigation2022}. Through a video grammar game called "\textit{Which English?}" it was shown that there is a "critical period" for learning a second language that extends into adolescence \cite{English2018}. And a dilemma game involving millions of people explored the moral values of our societies and how they vary between countries \cite{awad_moral_2018}.

Nevertheless, several challenges must be overcome to establish video games as a rigorous and reliable research tool in behavioural or social science \cite{gamesScience2023}. Synthetic data is a good candidate to address some of these challenges. For example, synthetic data can help with data privacy, fairness and augmentation, compensate for data deficiencies such as category imbalance or even produce data before the real one is available \cite{sytheticRoyalSociety}. Although synthetic data is not a replacement for real data, it can accelerate developments and reduce costs and effort. In recent years, interest in using synthetic data in social or behavioural science research has notably increased \cite{grund2022using} \cite{Quintana2020} \cite{SyntheticData2017}.

This paper's primary goal is to generate synthetic data for serious games based on interactive narratives, such as the one used in the RAYUELA project, which focuses on cyberbullying. To this end, we propose a simulator architecture and bring two innovations to the state of the art. First, we present a generic methodology to introduce external data (e.g., expert knowledge, survey data) to the simulator through probabilistic graphical models, particu\-larly Bayesian Networks (BN) \cite{pearl1988probabilistic}. Secondly, the synthe\-tic agent model is based on the Item Response Theory (IRT) cognitive modelling framework \cite{embretson2013item}. This latent variable framework has been extensively studied in the literature and has been shown to be far superior to classical test theory. In addition, it allows both the 'ability' of the participant and the usefulness or difficulty of each question to be inferred. To the best of our knowledge, IRT has only been used in the literature for statistical inference, never for generating synthetic data.

We have performed several Bayesian inference tests based on a hierarchical model to test the robustness and identifiability of the generated synthetic data. These experi\-ments test the identifiability of the experiment parameters and serve as a guide to the expected precision when using Bayesian inference with real player data.

\section{Simulator}
\subsection{Design Considerations}

Before detailing the proposed simulator architecture, we will review the design considerations and project constraints that led to the decisions made. Firstly, although the proposed architecture can be applied to other environments where participants must make a series of decisions or answer categorical questions, this work focuses on the specific problem of an interactive narrative serious game. Besides, it is noteworthy that in our work, the synthetic players do not aim to "win" the game but to approximate realistic human behaviours, in an approach more similar to \cite{wang_beyond_2019} and \cite{lin2019story}.

To ensure that the synthetic data better reflects reality, it is desirable to be able to introduce external information into the generative process (e.g. expert knowledge, surveys, prevalence data, etc.). It is also desirable to do this in a generic way, so that it is easy to experiment and introduce additional data at any time, and so that the proposed architec\-ture can be used to address issues other than cyberbullying. To meet this design need, we propose using probabilistic graphical models, particularly Bayesian Networks (BN) \cite{pearl1988probabilistic}. This model is a powerful visual and quantita\-tive tool for expressing probabilistic relationships among variables. BNs consist of a direct acyclic graph (DAG) structure that encodes which variables are causally related to others, and each node of the network contains a conditional probability table (CPT). The structure and parameters of a BN can be learned from data, manually constructed (usually with the help of experts in the specific problem being addressed), or a combination of both. In addition, a trained BN can be used to generate synthetic data by sampling from the learned probability distributions. 

The ultimate goal of RAYUELA's serious game is to identify different groups/clusters of players through the answers collected. In other words, to investigate whether the answers given in the serious game provide information about the players' behaviours in the real world. We can model this environment as a sequence of multi-choice questions, where each player's latent state changes the probability of choosing each option. This paper aims to generate synthetic data reflecting the internal states of the players and their cognitive decision-making process. Therefore, the proposed simulator must have a "decision maker" module that obtains the probabilities of choosing each answer from a given player profile and a question. To meet this design need, we propose using the IRT framework \cite{embretson2013item}, a testing theory based on the idea that the probability of a correct response to an item is a mathematical function of the respondent and item parameters. IRT is often regarded as superior to classical test theory \cite{embretson_item_2000}, primarily because in addition to inferring the "ability" of the participant, it also takes into account the "difficulty" of each question when assessing (and other possible parameters in complex models). Our work will use IRT to generate synthetic data rather than for statistical inference. With this approach, we achieve to model the interactions between players and the in-game decisions they have to confront based on widely used psychological theories. 

In summary, our proposed simulator models players' decision processes probabilistically using a widely used test theory (IRT), while incorporating expert knowledge and external data (e.g. surveys, prevalence data, etc.) through the use of BN.

\subsection{Architecture}

\begin{figure}[]
  \center
  \caption{Simulator's overall architecture and components. The generated \textit{agents} respond to the \textit{environment} and make a decision according to its \textit{profile} in a non-deterministic manner. The blue boxes represent external information that is fed into the simulation. The grey boxes represent the internal states and models of the synthetic agents.}
  \includegraphics[width=0.48\textwidth,keepaspectratio]{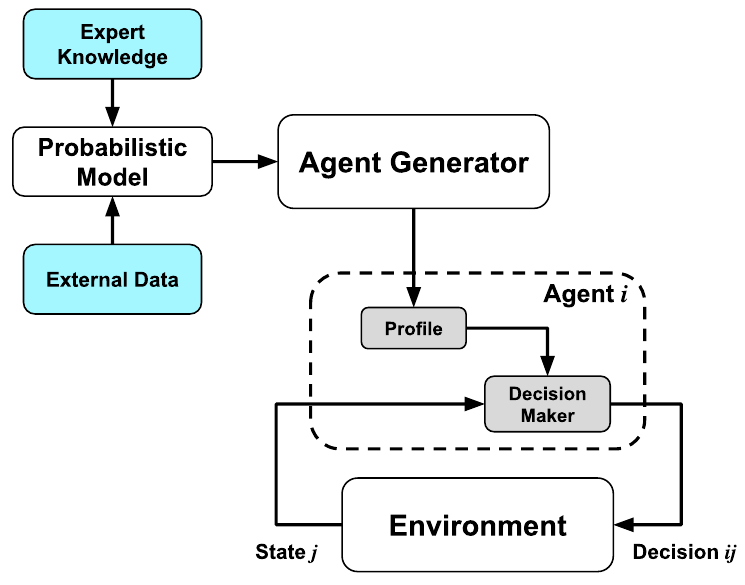}
  \label{fig:architecture}
\end{figure}

Considering the technical and design considerations outlined in the previous subsection, we summarise the proposed overall architecture of the simulator in Figure \ref{fig:architecture}. This modular architecture allows tweaking specific features of the simulator, thus avoiding future significant redesigns (e.g., if we want to create a new agent model, we would only have to modify that module). Four main components are found in this design:

\subsubsection{Probabilistic Model}
This module is responsible for incorporating expert knowledge and other external information (e.g., surveys or prevalence data) into the simulator, thus aligning the synthetic data with reality and making it as helpful as possible. This is achieved using a probabilistic model, such as BNs, where the expert knowledge is encoded into the network's DAG structure, and network parameters (CPTs) are learned from external information. External information can also be incorporated to define the prior belief probability distributions. Moreover, BNs allow us to \textit{interro\-gate} the model using "What if...?" questions to obtain quantifiable responses for events for which we have little or no data. 

We propose using a trained BN to generate synthetic data that the Agent Generator module will use to produce synthetic players with an individualised profile (in a proba\-bilistic way). In addition, the synthetic data generated by the BN will be incorporated into the final synthetic dataset to make it more informative and helpful. If we desire more control over the generation, we can condition chosen variables (e.g., Age=\textit{18}, Gender=\textit{Male}) to produce \textit{stratified} synthetic data. The proposed method to generate informed synthetic data using a BN can be summarised in the following steps:

\begin{enumerate}
    \item Build the BN structure (i.e., DAG) from expert know\-ledge, literature review or external data.
    \item Train the BN with external data to learn the parameters (i.e., CPTs) using a learning algorithm such as Maxi\-mum Likelihood Estimator or Expected Maximization \cite{EM_Algorithm_1977}.
    \item Obtain synthetic data from the BN using a sampling algorithm such as Bayesian Model Sampling or Gibbs sampling \cite{GibbsSampling1984}. The propagation of conditional probabilities will provide a probability distribution on the variable of interest (in our case study, having experienced a cyberbullying-related situation).
    \item Use the obtained probabilities in the variable of interest to determine the profile of each synthetic player/agent probabilistically.
\end{enumerate}

\subsubsection{Agent Generator}
This module generates synthetic agents with distinct parameters representing varied profiles (e.g., psychological or sociological profiles). The output of the Probabilistic Model (i.e., probabilities of the variable of interest) drives this generation process. Although the exact transformation process to obtain each synthetic player's profile is a design decision that will change drastically depending on the issue addressed and the number of profiles desired. It, therefore, allows for controlled generation at the service of researchers (e.g., generating an intentional imbalance in the synthetic data that more closely captures reality). Section \ref{sec:CaseStudy} will explain in detail the implementation we have done for our case study on cyberbullying.

\subsubsection{Agent}
This module aims to re-create the interaction of the synthetic agents with the simulator questions/dilemmas, obtaining as output the answers/decisions taken according to their profile (in a non-deterministic way). Two main components constitute the Agent module:

\begin{enumerate}[(i)]
    \item \textit{Profile} ($\alpha_i$): This is a fixed internal parameter, unique for each agent, representing the agent's profile. This numerical value is inherited from the Agent Generator module. For instance, in our case study on cyber\-bullying, the profile parameter will represent the risk propensity of each agent. Positive values of $\alpha_i$ would represent more risk-prone agents, and negative values represent agents with lower risk propensity. Values of $\alpha_i$ around zero represent a random player.
    \item \textit{Decision maker}: This submodule will simulate the decisions made by the agents in the game, according to the profile and question parameters, trying to align them probabilistically (thus capturing the uncertainty in human decision-making). The implementation is common to all agents.
\end{enumerate}

The approach implemented in the Decision Maker module borrows ideas from the IRT paradigm. However, some adjustments must be performed to make the model properly fit our particular case. As we explained before, the ultimate goal of our project's serious game is to identify different groups/clusters of players through the answers collected. Therefore, there will not be correct or incorrect answers, but answers representing greater alignment with certain profiles. 

In the simplest case, where agents will make dichotomous choices (i.e., two possible answers) can be formally expressed as Equation \eqref{eq:agent}. The answers of each player $i$ to each question $j$ are random samples from a Bernoulli probability distribution, with a probability $p_{ij}$ that depends on the agent's profile $\alpha_i\in \mathbb{R}$ and the question parameter $\beta_j\in [0,1]$, for $i \in [0, N]$ players and $j \in [0,Q]$ questions. Equation \eqref{eq:agent} is valid for dichotomous/binary questions, but it can be generalised to multiple choices questions by replacing the Bernoulli with a Categorical probability distribution and using a polytomous IRT-based model in the probability computation \cite{ostini2006polytomous}. 

\begin{equation}
        \text{Answer}_{ij} \sim \text{Bernoulli}(p_{ij}), \text{ with }
        p_{ij} = \frac{1}{1+e^{-\alpha_i\beta_j}} 
        \label{eq:agent}
\end{equation}

The question parameter $\beta_j$ is a numerical value unique for each question and represents its discriminatory ability to extract valuable information related to the agent's profile. It is inherited from the Environment module. A value of $\beta_j=0$ represents null information given by the question, and a value of $\beta_j=1$ represents perfect information. Namely, answering positively to a question with $\beta_j=1$ provides a good measurement of the agent's profile ($\alpha_i$). In actual games, a question with a value close to $\beta_j=0$ will represent those decisions whose answer is unrelated to the variable of interest (for instance, in our case study, it would be unrelated to cyberbullying).

\begin{figure}[!htp]
  \center
  \caption{Visualisation of the probability $p_{ij}$ equation values as a function of $\alpha$ and $\beta$ for the dichotomous/binary case. The values of the equation are shown for 7 values of $\alpha$, represented in different colours.}
  \includegraphics[width=0.49\textwidth,keepaspectratio]{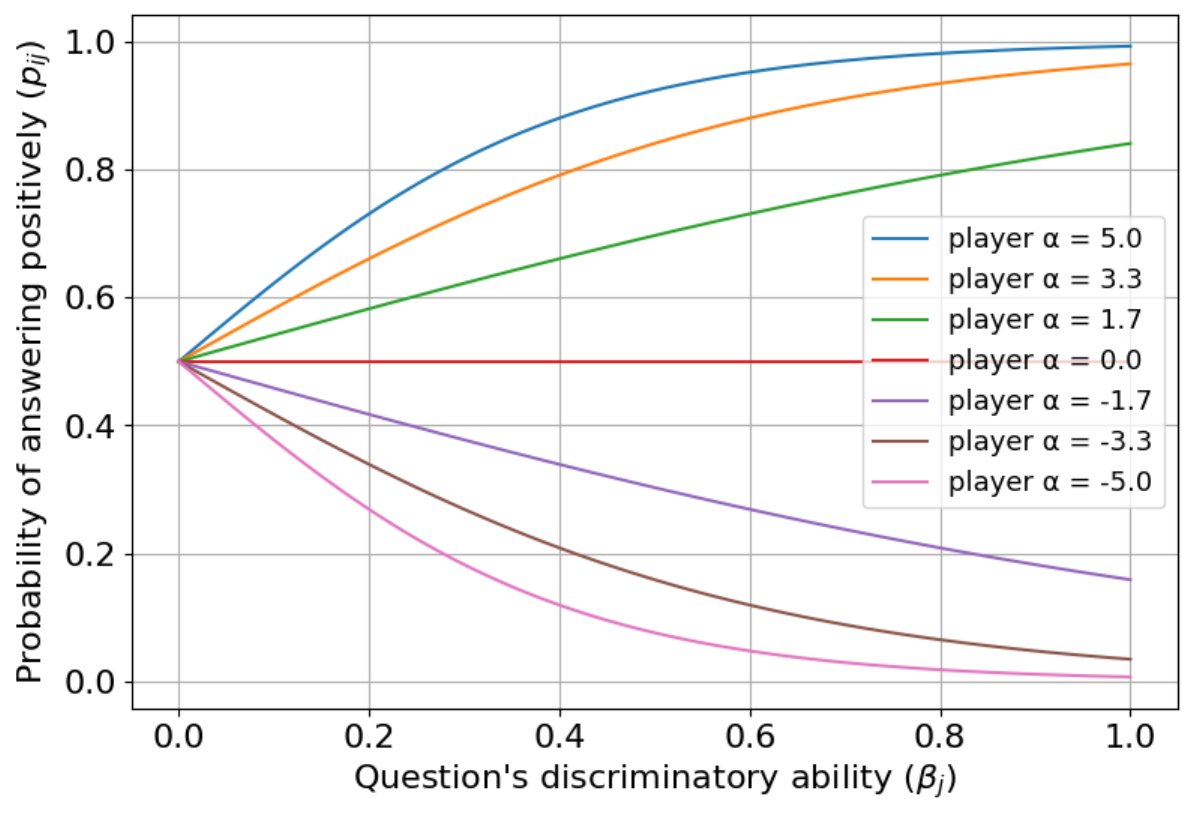}
  \label{fig:model_agent}
\end{figure}

Figure \ref{fig:model_agent} illustrates how the probability $p_{ij}$ of answering a question positively varies depending on the values of $\alpha$ and $\beta$. When $\beta\to 1$ (i.e., high discriminatory ability), agents have a high probability of choosing the response that matches their profile. However, when $\beta\to 0$ (i.e., low discriminatory ability), each agent has a probability that tends to $0.5$, regardless of the value of their individual $\alpha_i$. A random player ($\alpha_i=0$) will always answer randomly, regardless of the question or $\beta_j$ value.

\subsubsection{Environment}
This module simulates the game's narrative structure and is the component with which the synthetic agents interact. It provides the beta values of the questions to the agent module. In interactive narrative games, the internal structure of the scenarios and questions accessed by the player is in the form of a tree. Each node of the tree provides the possible choices that the agent can make in each question/situation of the game. As explained in the previous section, the $\beta_j$ parameter of the questions indicates its discriminatory ability to extract valuable information related to the agent's profile. During the simulation, these parameters are sampled from a probability distribution with values between 0 and 1 (e.g. Beta distribution).

\section{Case Study: Serious Game on Cyberbullying}
\label{sec:CaseStudy}

In this section we will explain how we applied the proposed simulator architecture to the serious game of the RAYUELA project \cite{RAYUELA}. The game is an interactive narrative focused on cyberbullying, aiming to identify different groups/clusters of players through the collected responses. Specifically, to differentiate between risky and safe players regarding their online behaviour.

Following the proposed architecture (Fig. \ref{fig:architecture}) and in order to generate synthetic data more faithful to reality, we will use a BN to introduce expert knowledge and external data into the simulation. The external data consists of a survey of minors in schools in Spain (Madrid and Valencia) during the year 2022. We collected 665 responses from students between 13 and 17 years old (Mean=14.5, SD=0.9), where 50.8\% identified themselves as males, 47.4\% as females, and 1.8\% as non-binary. In this survey, we collected a series of demographic data, some questions about the participants' relationship with new technologies (e.g., IoT devices) and the Internet, and finally, some inquiries about situations related to cyberbullying or cyber-harassment. Table \ref{table:sample_data_survey} shows a random sample of 5 survey participants.

\begin{figure}[]
  \center
  \caption{\textbf{Probabilistic Model}: Bayesian Network structure (DAG) which encodes the experts' hypotheses of causal relationships among the variables collected in the survey to minors.}
  \includegraphics[width=0.49\textwidth,keepaspectratio]{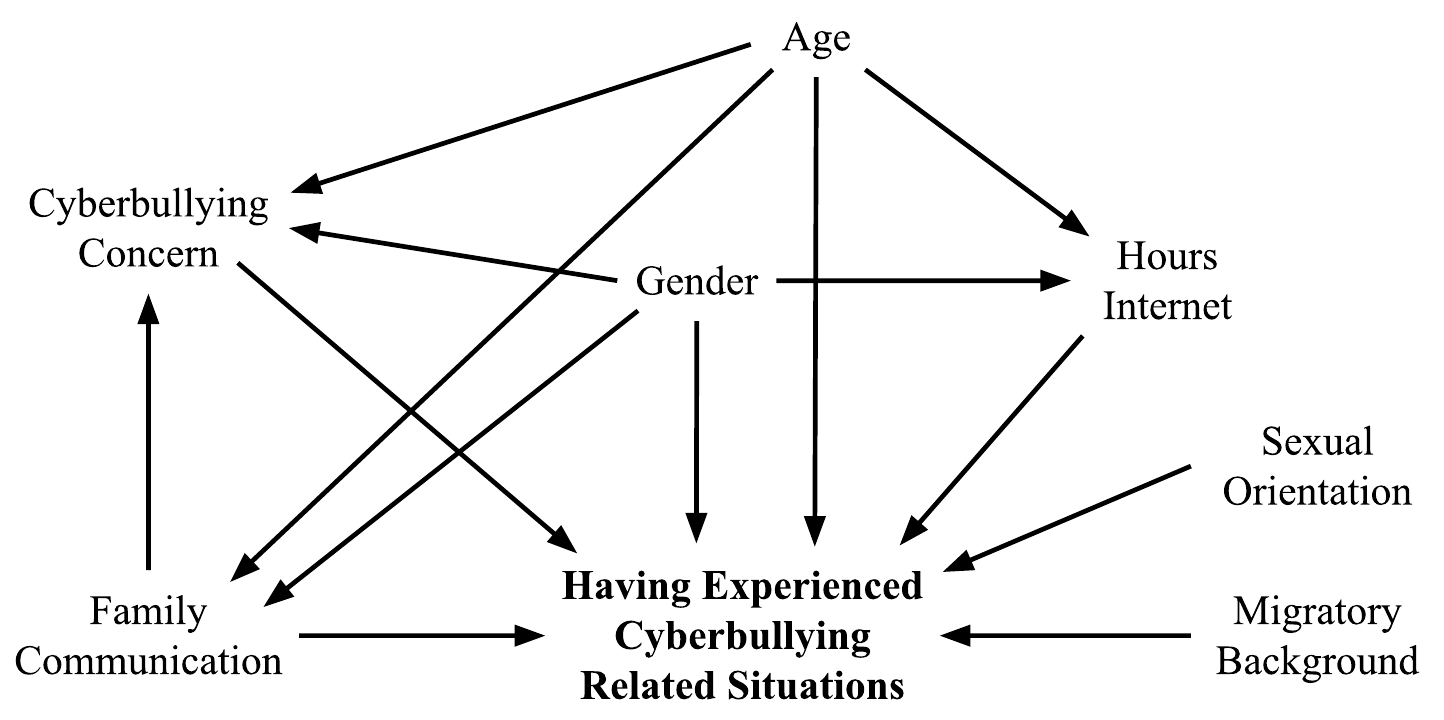}
  \label{fig:DAG}
\end{figure}

Using the variables collected through the survey and assisted by the group of psychologists with experience in cyberbullying, we have designed a structure for the BN (Figure \ref{fig:DAG}). This structure (DAG) encodes the experts' hypotheses of causal relationships among the variables. Afterwards, the parameters (CPTs) of the BN were trained using the data obtained from the survey and the Expected Maximization algorithm.

Once the BN has been trained, we begin to generate synthetic data using the Bayesian Model Sampling algorithm to finally obtain a binary probability distribution on the variable of interest (i.e., \textit{having experienced cyberbullying related situations}) for each synthetic agent. This probability on the variable of interest will condition whether the agent belongs to the risky or safe group/cluster and, therefore, the numerical value of its risk profile ($\alpha_i$).

\begin{table*}[!htp]
\caption{\textbf{External data}: Sample of 5 randomly selected survey participants. The last column, ''Having experienced situations related to cyberbullying in the last year'' is an aggregation of 3 questions in the survey about specific cyberbullying-related situations.}
\label{table:sample_data_survey}
\resizebox{\textwidth}{!}{%
\begin{tabular}{cccccccc}
\toprule
Gender & Age & \begin{tabular}[c]{@{}c@{}}Sexual\\ orientation\end{tabular} & \begin{tabular}[c]{@{}c@{}}Immigrant\\ background\end{tabular} & \begin{tabular}[c]{@{}c@{}}Daily hours on \\ the Internet\\ for leisure\end{tabular} & \begin{tabular}[c]{@{}c@{}}Cyberbullying\\ concern\\ (1-5)\end{tabular} & \begin{tabular}[c]{@{}c@{}}Family communication\\ on cyber-threats\\ (1-4)\end{tabular} & \begin{tabular}[c]{@{}c@{}}Having experienced situations\\ related to cyberbullying in the last year\\ (Aggregated)\end{tabular} \\
\midrule
Female & 13  & Bisexual                                                     & No                                                             & Between 3h and 4h                                                                    & 5/5 (very concerned)                                                      & 2/4 (rarely)                                                                            & No                                                                                                                               \\
Male   & 14  & Heterosexual                                                 & No                                                             & Between 1h and 2h                                                                    & 2/5 (unconcerned)                                                         & 3/4 (often)                                                                             & No                                                                                                                               \\
Male   & 16  & Bisexual                                                     & No                                                             & Between 3h and 4h                                                                    & 4/5 (concerned)                                                           & 3/4 (often)                                                                             & No                                                                                                                               \\
Female & 14  & Heterosexual                                                 & No                                                             & More than 4h                                                                         & 5/5 (very concerned)                                                      & 4/4 (very often)                                                                        & Yes                                                                                                                              \\
Female & 16  & Heterosexual                                                 & Yes                                                            & Between 3h and 4h                                                                    & 4/5 (concerned)                                                           & 3/4 (often)                                                                             & No                   \\
\bottomrule
\end{tabular}%
}
\end{table*}

For these two groups/clusters of players (risky and safe), we have made the assumption that their risk profiles ($\alpha_i$) are samples of two different Gaussian distributions. Allowing for some overlapping to account for the intrinsic uncertainty underlying human decision-making processes. In summary, the process to obtain the risk profile ($\alpha_i$) of each agent is as follows:

\begin{enumerate}
    \item Sample the probabilistic model (BN) to obtain synthe\-tic survey data and a probability of belonging to the risky/safe groups as a function of the other network variables (e.g. age, gender, immigrant background, etc.). Namely:

    $P_{\rm risky} = P(\text{risky}|\text{Bayesian Network})$

    $P_{\rm safe} = 1 - P_{\rm risky}$
    
    \item Use the obtained probabilities to decide probabilisti\-cally to which group each synthetic agent belongs. Namely:

    $
    G_i \sim \text{Bernoulli}(P_{\rm risky})
    $
    
    \item Once the group of each agent has been decided, sample the alpha value of the corresponding Gaussian distribution. Namely:

    $
    \begin{aligned}
        \alpha_i & \sim \begin{cases}\operatorname{Normal}\left(\mu_{\rm safe},\sigma_{\rm safe} \right) & \text { if } G_i=0 \\
        \operatorname{Normal}\left(\mu_{\rm risky},\sigma_{\rm risky} \right) & \text { if } G_i=1\end{cases}
    \end{aligned}
    $
    
\end{enumerate}

Once the BN has been trained with the survey data (i.e., external information) and we have defined the procedure for using the obtained probabilities to obtain the agents' risk profiles, we can start generating synthetic data. We have created a dataset of 500 synthetic players participating in a game simulation of 15 dichotomous/binary questions. For this case study, we have defined the hyperparameters for the Gaussian distributions as described in Equation \eqref{eq:distributions}. Note that the specific values we gave the hyperparameters are just an example that we believe is somewhat realistic. However, real players may behave differently, or the risky/safe groups may not even exist in the real world.

\begin{equation}
    \begin{array}{c}
    \alpha_i|\text{safe} \sim \text{Normal}(\mu=-2, \sigma=0.7)\\
    \alpha_i|\text{risky}\sim \text{Normal}(\mu=0.5, \sigma=1.2)
    \end{array}
    \label{eq:distributions}
\end{equation}

\begin{figure}[]
  \center
  \caption{Histogram of the risk profiles ($\alpha_i$) parameters of the synthetic generated dataset ($N=500$ agents). Lower $\alpha$ values encode agents with lower risk propensity and vice versa.}
  \includegraphics[width=0.47\textwidth,keepaspectratio]{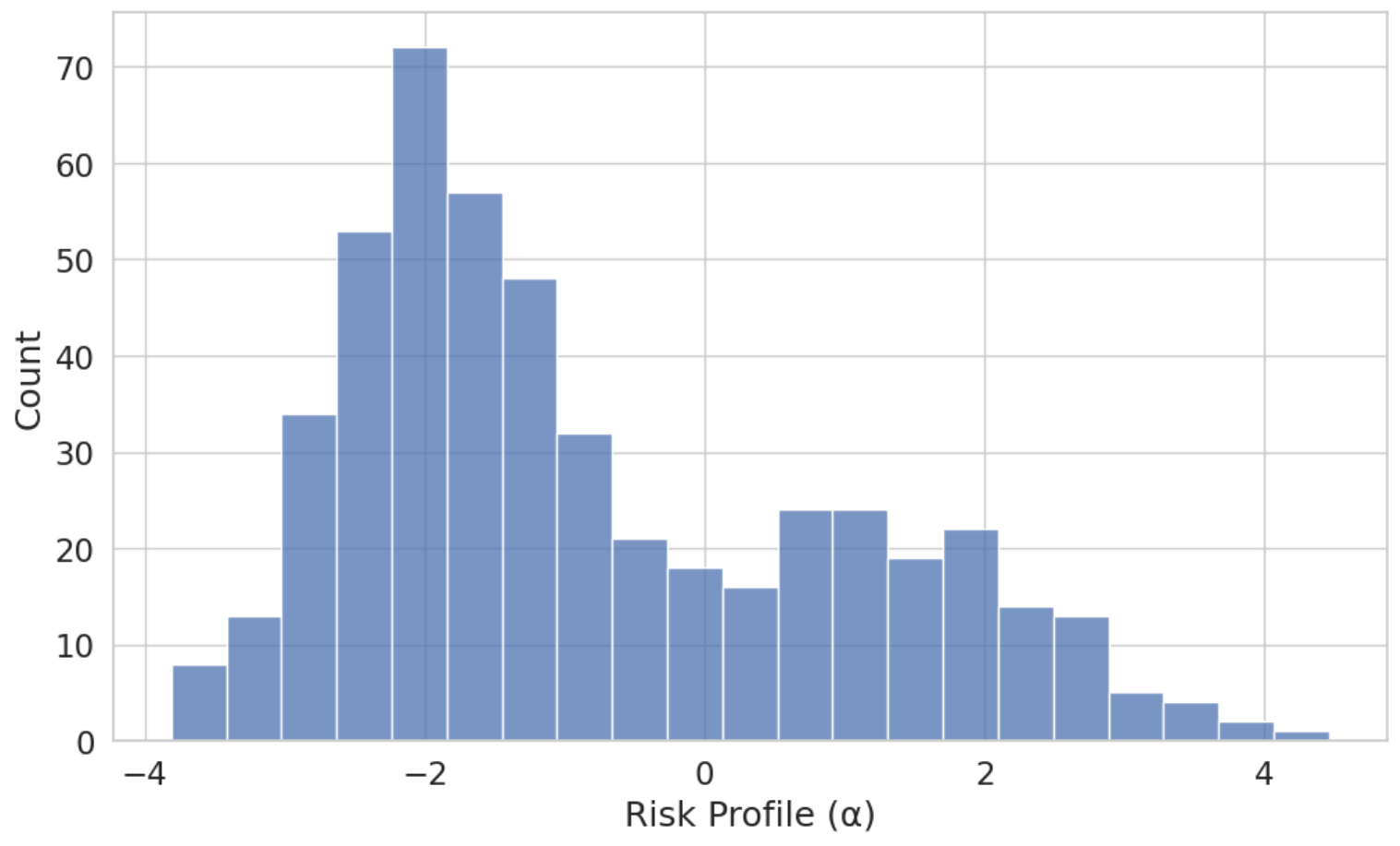}
  \label{fig:hist_risk}
\end{figure}

Table \ref{table:sample_data} shows 5 samples of the generated synthetic dataset, where each agent is stored in a row, and also includes synthetic personal information obtained from the BN (age, gender, sexual orientation, immigrant, hours of internet use, cyberbullying concern and family communication about cyber-threats). In the columns of the dataset where each agent's answers are stored (Q1 to Q15), the 1s mean that the agent chose the option implying the highest risk propensity. And the opposite with the 0s, the agent has chosen the option implying the lowest risk propensity. Figure \ref{fig:hist_risk} shows a histogram of the generated agents' risk-profile ($\alpha_i$) parameters. The bimodality reflects the fact that $\alpha_i$ comes from two different Normal distributions and the asymmetry because the incidence of risky profiles in the survey data is lower than 50\%.

\begin{table*}[!htp]
\caption{\textbf{Synthetic Data}: Sample of 5 agents randomly selected from the dataset generated ($N=500$ agents). Columns Q1 to Q15 indicate the questions of the simulation that has been created for this example. In those, the 1s means that the agent has chosen the option implying the highest risk propensity, and the 0s mean that it has chosen the option implying the lowest risk propensity.}
\label{table:sample_data}
\resizebox{\textwidth}{!}{
\begin{tabular}{ccccccccccccccc}
\toprule
\begin{tabular}[c]{@{}c@{}}Risk Profile\\($\alpha_i$)\end{tabular} & Q1 & Q2 & Q3 & ... & Q13 & Q14 & Q15 & Gender     & Age & \begin{tabular}[c]{@{}c@{}}Sexual\\ orientation\end{tabular} & \begin{tabular}[c]{@{}c@{}}Immigrant\\ background\end{tabular} & \begin{tabular}[c]{@{}c@{}}Daily hours on \\ the Internet\\ for leisure\end{tabular} & \begin{tabular}[c]{@{}c@{}}Awareness\\ Cyberbullying\\ (1-5)\end{tabular} & \begin{tabular}[c]{@{}c@{}}Family communication\\ on cyber-threats\\ (1-4)\end{tabular} \\
\midrule
-2.16                & 0  & 1  & 0  & ... & 0   & 0   & 0   & Male       & 13  & Heterosexual                                                 & No                                                             & Between 1h and 2h                                                                    & 4/5 (concerned)                                                           & 3/4 (often)                                                                             \\
1.69                 & 1  & 0  & 1  & ... & 0   & 1   & 0   & Female     & 16  & Heterosexual                                                 & No                                                             & More than 4h                                                                         & 2/5 (unconcerned)                                                         & 1/4 (never)                                                                             \\
0.42                 & 1  & 0  & 1  & ... & 1   & 1   & 0   & Female     & 14  & Heterosexual                                                 & No                                                             & Between 2h and 3h                                                                    & 4/5 (concerned)                                                           & 3/4 (often)                                                                             \\
-1.4                 & 1  & 0  & 0  & ... & 1   & 0   & 1   & Female     & 14  & Heterosexual                                                 & Yes                                                            & Between 2h and 3h                                                                    & 5/5 (very concerned)                                                      & 4/4 (very often)                                                                        \\
1.03                 & 0  & 0  & 0  & ... & 0   & 1   & 1   & Non-binary & 17  & Non-heterosexual                                             & No                                                             & Between 2h and 3h                                                                    & 5/5 (very concerned)                                                      & 1/4 (never)          \\
\bottomrule
\end{tabular}      
}      
\end{table*}

\subsection{Identifiability Analysis}
\begin{figure}[]
  \center
  \caption{Graphical representation of the hierarchical Bayesian model. Circular nodes represent continuous variables and square nodes discrete ones. Double-bordered nodes represent deterministic variables. Shaded nodes represent observed variables.}
  \includegraphics[width=0.45\textwidth,keepaspectratio]{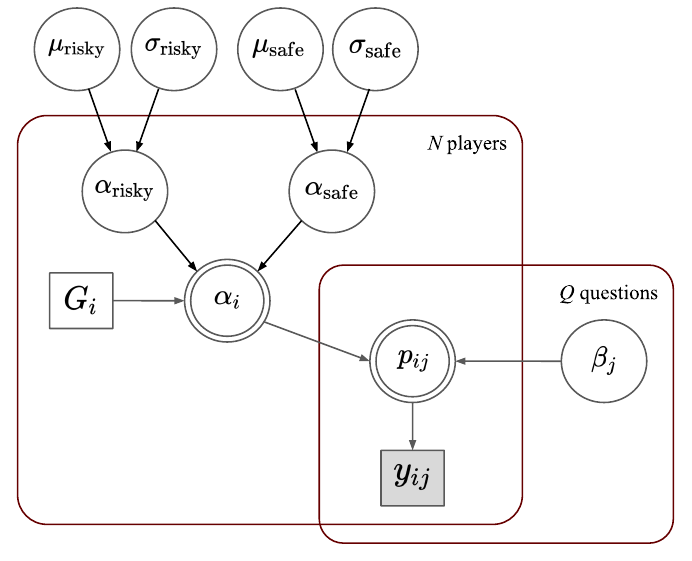}
  \label{fig:hierarchical}
\end{figure}

Once we have generated the synthetic data, we will perform an empirical identifiability analysis to ensure it is possible to estimate the parameters' values used in the simulator just from the generated data ($N=500$ players, $Q=15$ questions). Specifically, we will use a Bayesian hierarchical model with the same structure used to produce the data (described in the previous subsection) to estimate the hyperparameters 
defining the agents and the questions' $\beta_j$ parameters. We will use only the synthetic responses ($Q_1\ldots Q_{15}$) to train the Bayesian hierarchical model, as these are the data generated with the parameters we want to estimate. So, we will not use the synthetic data generated through the BN (e.g., gender, age, sexual orientation) to reconstruct these parameters. Equation \eqref{eq:priors} describes the prior distributions introduced in the hierarchical Bayesian model, and  Figure \ref{fig:hierarchical} shows the graphical representation. The parameter $p_{ij}$ is described in Equation \eqref{eq:agent}

In Bayesian inference, unlike in the Machine Learning or Deep Learning fields, we do not get a singular value due to the prediction; we obtain posterior probability distributions as a result. These distributions represent the epistemic uncertainty about the inferred statistical parameter condi\-tional on the collection of observed data. We have made the implementation using the open-source library \textit{PyMC} \cite{Salvatier2016}, a state-of-the-art software tool for probabilistic programming and statistical computation. 

\begin{equation}
    \begin{aligned}
        \mu_{\rm safe} & \sim \operatorname{Normal}(-1, 2) \\
        \sigma_{\rm safe} & \sim \operatorname{Exponential}(1) \\
        \mu_{\rm risky} & \sim \operatorname{Normal}(1, 2) \\
        \sigma_{\rm risky} & \sim \operatorname{Exponential}(1) \\
        G_i & \sim \operatorname{Bernoulli}(0.5) \\
        \alpha_{i} & \leftarrow \begin{cases}\operatorname{Normal}(\mu_{\rm safe}, \sigma_{\rm safe}) \text { } \text { } \text { } \text { if } G_i=0 \\
        \operatorname{Normal}(\mu_{\rm risky}, \sigma_{\rm risky}) \text { if } G_i=1\end{cases} \\
        \beta_{j} & \sim \operatorname{Beta}(1,1)\\
        y_{ij} & \sim \operatorname{Bernoulli}(p_{ij})
    \end{aligned}
    \label{eq:priors}
\end{equation}

\begin{figure}[]
  \center
  \caption{Posterior probability distributions of the hyperparameters of the Gaussian distributions generating the agents' profiles in the case study. The true values used in the generation process are shown in orange. The black line at the bottom of each plot represents the HDI (94\%).}
  \includegraphics[width=0.35\textwidth,keepaspectratio]{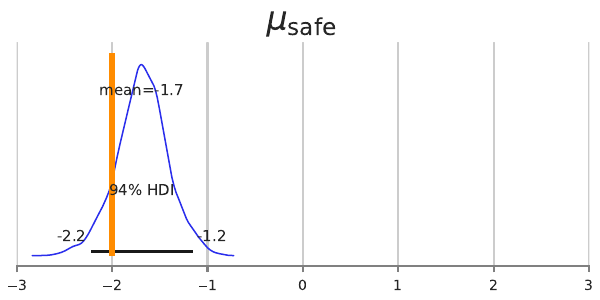}
  \includegraphics[width=0.35\textwidth,keepaspectratio]{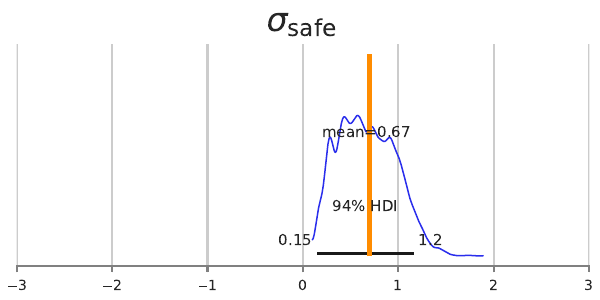}
  \includegraphics[width=0.35\textwidth,keepaspectratio]{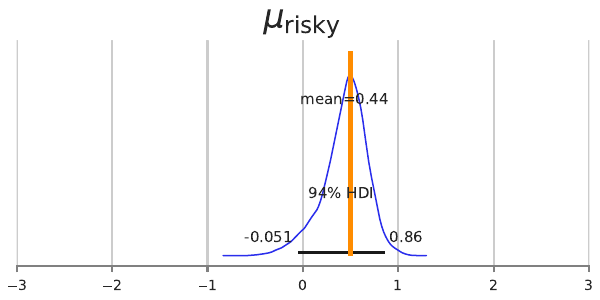}
  \includegraphics[width=0.35\textwidth,keepaspectratio]{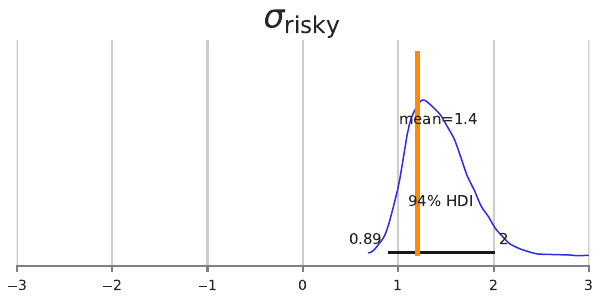}
  \label{fig:hyper_posteriors}
\end{figure}

\begin{figure}[]
  \center
  \caption{HDIs (94\%) of the posterior probability distributions of the questions' parameters ($\beta_j$). The true values used in the generation process are shown in orange. Note that all the HDIs of the inferred parameters, except $\beta_3$, include the true value.}
  \includegraphics[width=0.45\textwidth,keepaspectratio]{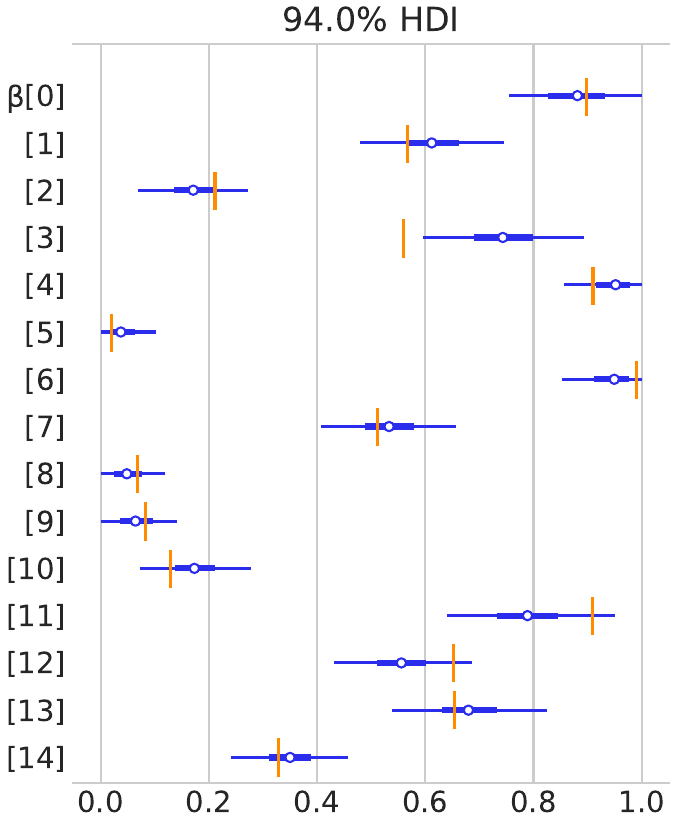}
  \label{fig:betas}
\end{figure}

In Figure \ref{fig:hyper_posteriors}, we find the posterior probability distributions of the hyperparameters of the Gaussian distributions that generate the agents' profiles. Figure \ref{fig:betas} shows the posteriors of the beta parameters of the questions. In both figures, the true value used in the generation process is shown in orange. Both figures also show the High Density Interval (HDI) of the posterior distributions \cite{mcelreath2020statistical}. The HDI is an interval within which the value of an unobserved parameter falls with a certain probability. In Bayesian inference, if the true parameter is within the 94\% HDI of the posterior distribution, it is usually considered a "correct guess" \cite{Makowski2019}.

As can be seen, the parameters were reconstructed quite accurately for the setting presented in this analysis. Therefore, in this sense, the synthetic data generation was successful, as the generated data encapsulate (probabilistic) information about the agent groups/clusters and the discri\-minative ability of the questions.

\subsection{Robustness Analysis}
Following the analysis of the synthetic data and the proposed model, in this subsection we will analyse the robustness in the reconstruction of the parameters using the synthetic data, as a function of the number of agents and questions. To do so, we have used the hierarchical Bayesian model shown in Figure \ref{fig:hierarchical}.

The motivation for this analysis is that, as we have seen in the previous subsection, the parameters used to generate the synthetic data are reconstructable (i.e. the true value is in the 94\% HDI of the posteriors). However, if these posteriors are too wide (i.e., low confidence in the prediction), they will not be helpful, even if the true value is still within the HDI range. Therefore, in this analysis, we will systematically examine the "width" of the estimated posteriors while varying the number of agents and questions generated. In other words, we will analyse the confidence with which the Bayesian model has inferred the generation parameters.

To quantify the "width" of the posterior distributions with a single metric, we will use the entropy \cite{Shannon_Entropy} of the distributions. This metric measures the average amount of \textit{information} or \textit{uncertainty} in a random variable. Given a discrete random variable $X$, which takes values in the range of $\mathcal{X}$ and is distributed according to the probabilities $p$, Equation \ref{eq:entropy} defines its entropy. 

\begin{equation}
H(X)=-\sum_{x \in \mathcal{X}} p(x) \log_2 p(x)=\mathbb{E}[-\log_2 p(X)]
\label{eq:entropy}
\end{equation}

\begin{figure}[]
  \center
  \caption{Robustness experiments using the hierarchical Bayesian model on the agents' parameters ($\alpha_i$), varying the number of agents and questions. The results show the normalised entropy of the posterior distribution of the inferred parameters, being 1 maximum entropy (i.e., no valuable information about the parameters) and 0 minimum entropy (i.e., complete information about the true value of the parameters).}
  \includegraphics[width=0.48\textwidth,keepaspectratio]{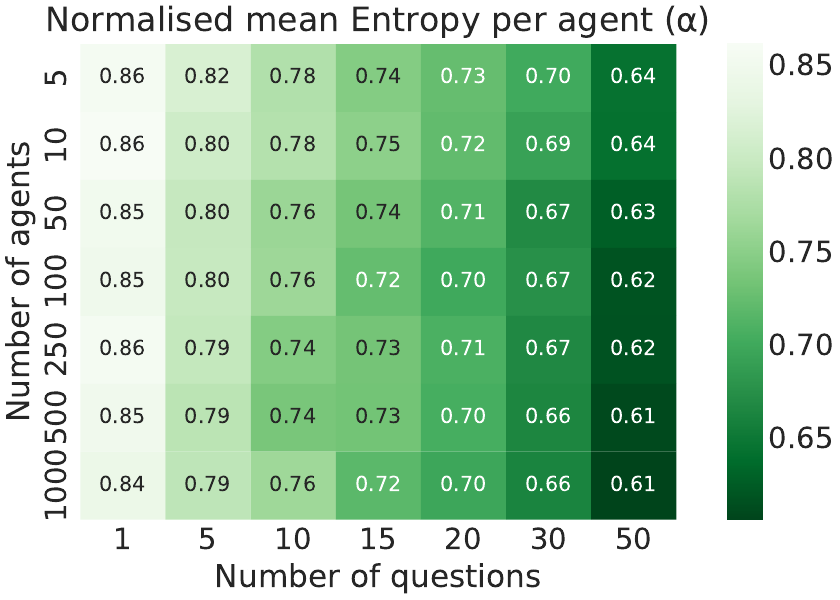}
  \label{fig:entropy_alpha}
\end{figure}

\begin{figure}[]
  \center
  \caption{Robustness experiments using the hierarchical Bayesian model on the questions' parameters ($\beta_j$), varying the number of agents and questions. The results show the normalised entropy of the posterior distribution of the inferred parameters, being 1 maximum entropy (i.e., no valuable information about the parameters) and 0 minimum entropy (i.e., complete information about the true value of the parameters).}
  \includegraphics[width=0.48\textwidth,keepaspectratio]{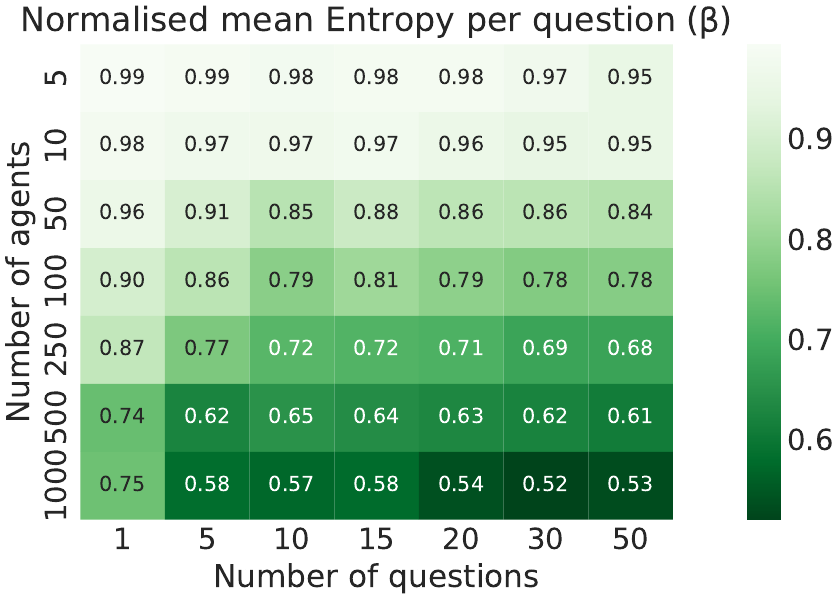}
  \label{fig:entropy_beta}
\end{figure}

In the experiments, we varied the number of agents from 5 to 1000 and the number of questions from 1 to 50. Then, we trained the hierarchical Bayesian model on each combination. Subsequently, we calculated the average entropy of $P(\alpha_i|\text{Data})$ and $P(\beta_j|\text{Data})$. Low entropy values represent that the model has high confidence in its prediction (i.e., the distribution is narrow) and vice versa. As we treat $p(x)$ as discrete (sample) probabilities, and to be able to compare among sets of parameters,  we make a histogram of each distribution with the same number of bins in the same range of the parameter. To reduce sampling variability, we performed each experiment (with a fixed number of players and questions) 5 times, then normalised and averaged the obtained entropies. The final results are shown in Figs. \ref{fig:entropy_alpha} and \ref{fig:entropy_beta}. 

\begin{figure}[]
  \center
  \caption{Examples of posterior probability distributions of the $\alpha$ parameters. The corresponding normalised entropy value is also indicated to facilitate the interpretation of the results obtained in the robustness experiments.}
  \includegraphics[width=0.35\textwidth,keepaspectratio]{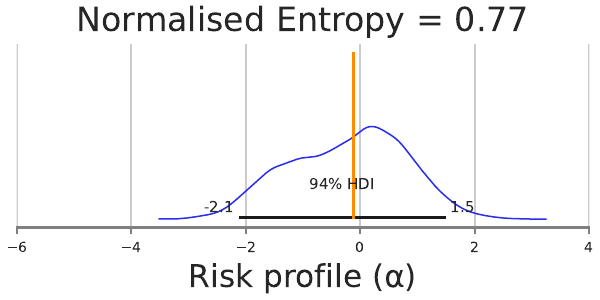}
  \includegraphics[width=0.35\textwidth,keepaspectratio]{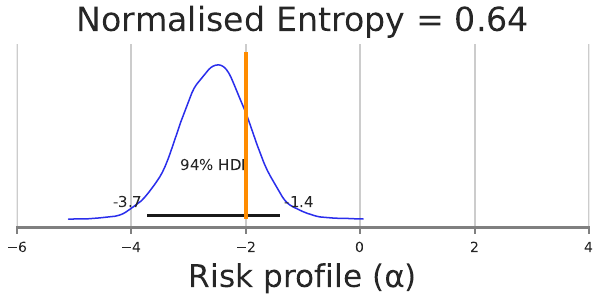}
  \label{fig:posterior_examples}
\end{figure}

The entropy value 1 represents complete uncertainty (i.e., the data do not contain any information about the parameter), and 0 represents perfect parameter information. To further facilitate the interpretation of the results obtained in the heatmaps, in Figure \ref{fig:posterior_examples} we show two examples of posterior distributions of the $\alpha$ parameters, also indicating the corresponding normalised entropy value.

The results of this robustness analysis give us an indica\-tion of how many players or questions we will need, depen\-ding on the precision with which we want to estimate the latent parameters. If we assume that the proposed model reflects the behaviour of real players sufficiently well, this will help us to design the serious game of the RAYUELA project (our case study) and give us an idea of the precision we can expect depending on the number of participants, thus speeding up the development process.

\section{Conclusions}
In this work, we have proposed and implemented a simulator capable of generating probabilistic synthetic data for serious games based on interactive narratives. For this purpose, we have created a generic modular simulator (Fig. \ref{fig:architecture}) that allows tackling similar problems with little or no redesign effort. The proposed architecture makes two main contributions to state of the art. On the one hand, we propose a methodology to introduce external data (e.g., expert knowledge, survey data) into the simulation through probabilistic graphical models, particularly Bayesian net\-works. On the other hand, the mathematical modelling of the synthetic agents is based on the Item Response Theory (IRT) framework. This latent variable framework has been extensively studied in the literature and has been shown to be far superior to classical test theory. To the best of our knowledge, IRT has only been used in the literature for statistical inference, never for generating synthetic data. 

Synthetic data generation offers several benefits, such as augmenting data, dealing with missing data or improving fairness and privacy. In our case, one clear-cut benefit is that creating unlimited but also variable meaningful data allows us to anticipate the data modelling and analysis, speed up the development process and reduce the cost of our use case.
The use case presented is framed in the RAYUELA project \cite{RAYUELA}, where we intend to use a serious game based on an interactive narrative to prevent and better understand the phenomenon of cyberbullying. Using the data from a custom-built survey of minors in Spain, we built our model using the proposed architecture, expert knowledge, and external data. In particular, we collected 665 responses from students between 13 and 17 years old in this survey. The survey gathers demographic data, questions about the participants' relationship with new technologies and the Internet, and inquiries about cyberbullying or cyber-harassment.

We have performed several Bayesian inference tests based on a hierarchical model to test the robustness and identifiability of the generated synthetic data. The Bayesian nature of the model allows us to infer the model's confidence (uncertainty) in its predictions. These experiments test the identifiability of the experiment parameters and serve as a guide to the expected precision when using Bayesian inference with real player data. Furthermore, using the IRT paradigm with Bayesian inference allows us to infer parameters about the game questions and deduce their discriminatory ability (in our case study, their ability to differentiate players according to their risk propensity).

With this paper, we aim to fill the gap in the literature on synthetic data generation in serious games and thus help other researchers to address typical problems such as data sparsity or privacy and to accelerate development processes. Among the many possible future lines of work, we can highlight the validation of the proposed IRT model with real players answering questions in the RAYUELA game. 

\section*{Acknowledgement}
This work has received funding from the European
Union’s Horizon 2020 research and innovation programme
under grant agreement No 882828. The authors would like to
thank all the partners within the consortium for the fruitful
collaboration and discussion. The sole responsibility for the
content of this document lies with the authors and in no way
reflects the views of the European Union



\bibliographystyle{ieeetr}
\bibliography{cas-dc-template.bib}

\end{document}